\documentclass[aps, prl,twocolumn,showpacs,superscriptaddress]{revtex4}

\usepackage{graphicx}

\usepackage{ulem}
\usepackage{color}
\usepackage{amssymb}

\def\journal#1#2#3#4#5{#1, #2 {\bf #3}, #4 (#5).}

\def\PRB{Phys.\ Rev.\ B}

\newcommand{\chiSG}{\ensuremath{\chi_\mathrm{SG}}}
\newcommand{\TSG}{\ensuremath{T_\mathrm{f}}}
\newcommand{\Tc}{\ensuremath{T_\mathrm{c}}}
\newcommand{\TN}{\ensuremath{T_\mathrm{N}}}
\newcommand{\Ns}{\ensuremath{N_\mathrm{s}}}
\newcommand{\qEA}{\ensuremath{q_\mathrm{EA}^2}}

\definecolor{green}{rgb}{0.0, 0.5, 0.0}

\begin{document}
\title{
Spin-glass Transition in Bond-disordered Heisenberg Antiferromagnets 
Coupled with Local Lattice Distortions on a Pyrochlore Lattice 
}
\author{Hiroshi Shinaoka}
\altaffiliation[Present address: ]{NRI, AIST, Tsukuba 305-8568, Japan}
\affiliation{Institute for Solid State Physics, University of Tokyo, Kashiwanoha, Kashiwa, Chiba, 277-8581, Japan}
\author{Yusuke Tomita} 
\affiliation{Institute for Solid State Physics, University of Tokyo, Kashiwanoha, Kashiwa, Chiba, 277-8581, Japan}
\author{Yukitoshi Motome} 
\affiliation{Department of Applied Physics, University of Tokyo, 7-3-1 Hongo, Bunkyo-ku, Tokyo 113-8656, Japan}
\date{\today}

\begin{abstract}
Motivated by puzzling characteristics of spin-glass transitions widely observed in pyrochlore-based frustrated materials, we investigate effects of coupling to local lattice distortions in a bond-disordered antiferromagnet on the pyrochlore lattice by extensive Monte Carlo simulations.
We show that the spin-glass transition temperature $\TSG$ is largely enhanced by the spin-lattice coupling, and furthermore, becomes almost independent of $\Delta$ in a wide range of the disorder strength $\Delta$.
The critical property of the spin glass transition is indistinguishable from that of the canonical Heisenberg spin glass in the entire range of $\Delta$.
These peculiar behaviors are ascribed to a modification of the degenerate manifold from continuous to semidiscrete one by the spin-lattice coupling.
\end{abstract}


\maketitle

In the last few decades, increasing attention has been devoted to low-temperature($T$) behavior of geometrically frustrated magnets~\cite{Diep05}.
Frustration suppresses conventional long-range ordering such as N\'{e}el ordering down to much lower $T$ compared to the interaction energy scale, opening the possibility of alternative low-$T$ phases.
The spin glass (SG), in which spins are frozen randomly, is one of such possibilities widely observed in geometrically frustrated materials~\cite{Ramirez90, Martinho01, Gardner10, Tristan05, Munekata06, Zhou08}.
However, it is unclear so far how the nature of SG is different from the canonical one driven solely by randomness~\cite{CanonicalSG}.

\begin{figure}[!]
 \centering
 \includegraphics[width=.425\textwidth,clip]{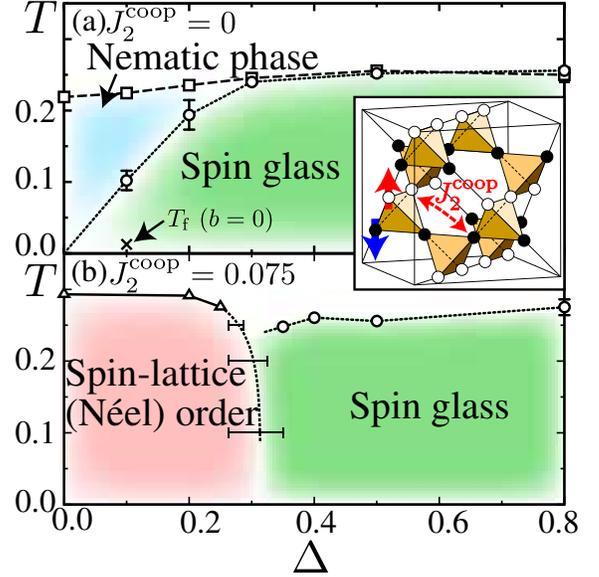}
 \caption{(color online). $\Delta$-$T$ phase diagrams obtained by MC simulation with (a) $J_2^\mathrm{coop}=0.0$ and (b) $0.075$ for $b=0.2$. The nematic ($\Tc$), antiferromagnetic ($\TN$), and SG transition temperatures ($\TSG$) are denoted by squares, triangles, and circles, respectively.
 In (a), $\TSG$ coincides with $\Tc$ for $\Delta \gtrsim 0.3$, suggesting a multicritical point at $\Delta \simeq 0.3~(\simeq b)$.
 The cross in (a) denotes $\TSG$ for $b=0$ and $\Delta=0.1$~\cite{Tam10}.
 The inset shows a 16-site cubic unit cell of the pyrochlore lattice. The open and filled circles in the inset denote two nonequivalent sites with opposite spins in the N\'{e}el order. 
 The N\'{e}el-SG phase boundary in (b) was determined by the inflection point of $m_{\text{s}}^2(\Delta)$ curve for $L=4$ [see the inset of Fig.~2(f)].
}
 \label{fig:pd}
\end{figure}

Antiferromagnets on a pyrochlore lattice (the inset of Fig.~\ref{fig:pd}) are typical examples of geometrically frustrated spin systems.
When considering classical Heisenberg spins with nearest-neighbor exchange interactions, no long-range ordering occurs down to zero $T$, and the ground state has continuous macroscopic degeneracy~\cite{Reimers92, Moessner98a}.
Recently, the effect of randomness in the exchange interactions was studied upon this degenerate manifold~\cite{Bellier-Castella01, Saunders07, Andreanov10, Tam10}. 
It was shown that the randomness immediately lifts the degeneracy, inducing a SG transition:   
The transition temperature $\TSG$ is proportional to the disorder strength $\Delta$. 
This gives a clue to explain why SG is prevailing in geometrically 
frustrated materials.

However, several characteristics of the SG still remain puzzling.
One of the surprising aspects is that, in many pyrochlore-based magnets, $\TSG$ appears to be almost independent of the strength of disorder $\Delta$. For example, for a typical frustrated SG compound Y$_2$Mo$_2$O$_7$ with $\TSG\simeq 22\text{K}$~\cite{Greedan86, Raju92, Gingras96, Gingras97, Miyoshi00}, 
random substitution of Y$^{2+}$ by La$^{2+}$ does not change $\TSG$ for the La concentration up to 50\%, despite a substantial change of the Curie-Weiss temperature $\theta_\mathrm{CW}$~\cite{Sato87}. Similar behavior was observed also in spinel oxides (Zn$_{1-x}$Cd$_x$)Cr$_2$O$_4$ for $x \le 0.1$, in which SG emerges after the antiferromagnetic spin-lattice order at $x=0$ is destroyed by small Cd substitution~\cite{Ratcliff02}.
Another distinctive aspect is that $\TSG$ is much higher than 
that theoretically expected for a moderate strength of disorder $\Delta$; 
e.g., a numerical estimate of $\TSG/J\simeq 0.01$ for $\Delta/J=0.1$ is about 20--30 times smaller than 
the experimental value~\cite{Saunders07, Andreanov10,Tam10}. 
These behaviors indicate that $\TSG$ is not set by $\Delta$, and suggest that some important factor is missing in the previous theories~\cite{Bellier-Castella01, Saunders07, Andreanov10,Tam10}.

A candidate is the magnetoelastic coupling.
Importance of local lattice distortions has been pointed out for Y$_2$Mo$_2$O$_7$ by various microscopic probes~\cite{Booth00,Keren01,Sagi05,Greedan09,Ofer10}.
They are crucial also in (Zn$_{1-x}$Cd$_x$)Cr$_2$O$_4$, as obviously seen in the spin-lattice ordering at $x=0$~\cite{Ratcliff02}.
Theoretically, it was shown that the bond randomness destroys the spin-lattice ordering and induces a SG state~\cite{Saunders08}.
However, the argument was limited to a uniform global lattice distortion, and $\TSG$ is deduced to behave similarly to the case in the absence of the spin-lattice coupling, i.e., $\TSG\propto \Delta$. 
Hence, the puzzling behaviors of $\TSG$ still remain unresolved.

In this Letter, we investigate the effect of coupling to local lattice distortions on the SG transition in a bond-disordered classical Heisenberg antiferromagnet on the pyrochlore lattice.
We show that $\TSG$ is largely enhanced by the spin-lattice coupling $b$, and pinned at $T\simeq b$, i.e., almost independent of $\Delta$ in a wide regime of $\Delta$.
We clarify the mechanism of the peculiar behaviors 
by considering how the spin-lattice coupling affects the degenerate manifold in the model.

We take as a starting point the Hamiltonian
\begin{equation}
	\mathcal{H}= \sum_{\langle i,j \rangle} 
	\Big[ J_{ij} \left(1-\alpha \rho_{ij}\right)\vec{S}_i \cdot \vec{S}_j + \frac{K}{2}\rho_{ij}^2 
	\Big],\label{eq:bigHam}
\end{equation}
where the sum runs over the nearest neighbor bonds of the pyrochlore lattice, and $\rho_{ij}$ is the change in distance between neighboring Heisenberg spins $\vec{S}_i$ and $\vec{S}_j$, relative to the equilibrium lattice constant.
The model describes both the exchange randomness in $J_{ij}$ induced by static bond disorder, and magnetoelastic coupling to local lattice distortions $\rho_{ij}$.
Here we introduce a uniformly-distributed randomness as $J_{ij} \in [J-\Delta,J+\Delta]$ with $0\le \Delta < J$. 
The last term in Eq.~(\ref{eq:bigHam}) represents the elastic energy of lattice distortions. 
Hereafter, all the energy scales including $T$ are measured in units of $J$.

In general, the lattice distortions $\rho_{ij}$ depend on each other; that is, they have a cooperative aspect, which may lead to structural transition concomitant with some magnetic ordering. 
When the cooperative aspect is ignored, the model (\ref{eq:bigHam}) is much simplified by integrating out $\rho_{ij}$. The resultant spin-only Hamiltonian is given by
\begin{equation}
	\mathcal{H}= \sum_{\langle i,j \rangle} 
	\Big[ J_{ij} \vec{S}_i \cdot \vec{S}_j - b_{ij} 
	\big( \vec{S}_i \cdot \vec{S}_j 
	\big)^2 
	\Big], \label{eq:Ham}
\end{equation}
where $b_{ij}$ is the biquadratic coupling given by $J_{ij}^2 \alpha^2/2K >0$, 
also being bond-disordered variables.
Hereafter we use $b\equiv\alpha^2/2K$ to measure the strength of the spin-lattice coupling.
We consider the model (\ref{eq:Ham}) to unveil intrinsic effects of $b$. 
The result will be discussed for Mo pyrochlores that show no spin-lattice ordering. 
To discuss the case in which a spin-lattice ordering appears as in (Zn$_{1-x}$Cd$_x$)Cr$_2$O$_4$, 
we take the cooperative aspect into account in a form of effective spin-spin interactions, following the results in previous theoretical studies~\cite{Tchernyshyov02}.
Among many possible contributions, we focus on the simplest one from anti-correlations between neighboring bond distortions by introducing a next-nearest neighbor term $\mathcal{H}_\mathrm{coop} \equiv  J_2^\mathrm{coop} \sum_{\langle\langle i,j \rangle\rangle} \vec{S}_i \cdot \vec{S}_j$ to Eq.~(\ref{eq:Ham}) with $J_2^\mathrm{coop}>0$ (the sum is over the second neighbor pairs; see the inset of Fig.~\ref{fig:pd})~\footnote{We consider the spatially-uniform $J_2^\mathrm{coop}$ as a control parameter for a spin-lattice ordering. A randomness in $J_2^\mathrm{coop}$ will reduce the N\'{e}el state but hardly affect the SG behavior.}.
A similar term is derived from the coupling to the site phonon~\cite{Bergman06}.

When $b$ and $J_2^\mathrm{coop}$ are both zero, which we call the bilinear limit, the model reduces to the one previously studied~\cite{Bellier-Castella01, Saunders07, Andreanov10,Tam10}. 
For a finite spin-lattice coupling $b>0$ but $J_2^\mathrm{coop}=0$, the present model at $\Delta=0$ exhibits a weak first-order nematic transition at $\Tc \sim b$, below which spins select a common axis without selecting their directions on it.
The ground state is identified by a set of spin-ice type local constraints that two of four spins are opposite to the other two on every tetrahedron.
This leaves the system magnetically disordered down to $T=0$ with the \textit{semi-discrete} degeneracy~\cite{Shannon10}.
$J_2^\mathrm{coop}$ lifts the degeneracy and stabilizes a spin-lattice (N\'{e}el) order over the nematic phase. 
The antiferromagnetic $J_2^\mathrm{coop}$ induces the ${\bf q}={\bf 0}$ four-sublattice N\'{e}el order~\cite{Chern08}, schematically shown in the inset of Fig.~\ref{fig:pd}.
Our interest here is how the SG transition is induced by $\Delta$ in the competition with the nematic and N\'{e}el orderings. 

To address the issue, we employ an extension of the loop algorithm which enables an ergodic sampling over the semidiscrete degenerate manifold at low $T \ll b$~\cite{Shinaoka10a}. 
We also adopt the exchange MC method~\cite{Hukushima96} and the overrelaxation update~\cite{Alonso96}.
We consider periodic systems of cubic geometry with $L^3$ unit cells with totally $\Ns = 16L^3$ spins.
To identify the SG, nematic, and antiferromagnetic transitions, we calculate the SG susceptibility $\chiSG$, nematic order parameter $Q^2$, sublattice magnetization $m_{\text{s}}$, and specific heat $C$.
$\chiSG$ is given by $\Ns \qEA$, where $\qEA$ is the Edwards-Anderson order parameter defined by $\qEA \equiv \Ns^{-1} \langle \langle \sum_{\mu,\nu=x,y,z} (\sum_{i=1}^{N_\mathrm{s}} S_{i\mu}^\alpha S_{i\nu}^\beta)^2 \rangle_T \rangle_\Delta$~\cite{Edwards75}. 
Here $\langle \cdots \rangle_T$ denotes a thermal average and 
$\langle \cdots \rangle_\Delta$ a random average over the interaction sets; 
the upper suffixes $\alpha$ and $\beta$ denote two independent replicas of the system with the same interaction set.
The nematic order parameter $Q^2$, which measures the spin collinearity, is defined as $Q^2 \equiv 2 \Ns^{-2} \langle \langle \sum_{ij} \{ (\vec{S}_i \cdot \vec{S}_j)^2 - 1/3 \} \rangle_T \rangle_\Delta$.
The sublattice magnetization is defined as $m_{\text{s}}\equiv 4 (\sum_{l} |\langle \sum_{i \in l} \vec{S}_i\rangle|^2 / \Ns^2)^{1/2}$, where $l$ labels the four sublattices of the pyrochlore lattice~\cite{Chern08}.
The specific heat $C$ is calculated by $C=\Ns^{-1} T^{-1} \langle \langle \mathcal{H}^2 \rangle_T - \langle \mathcal{H} \rangle_T^2 \rangle_\Delta$.
All data shown below are averaged over a number of interaction sets varying from $2000$ to $100$, and typical MC steps for measurement vary from $10^4$ to $10^7$ depending on $L$ and $\Delta$.

\begin{figure}[t]
 \centering
 \includegraphics[width=0.5\textwidth,clip]{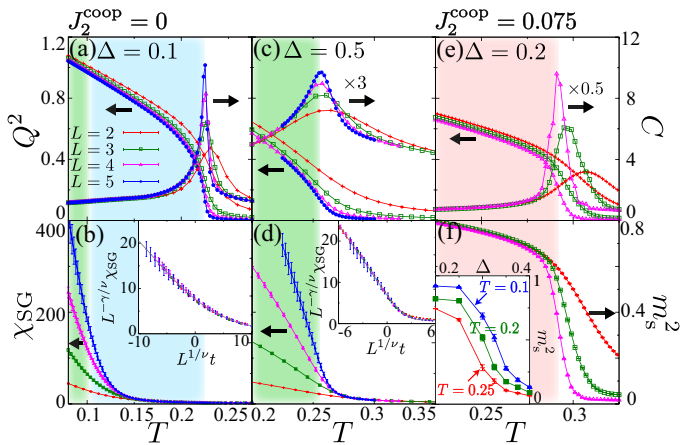}
 \caption{(color online). Temperature dependences of the nematic order parameter $Q^2$,  specific heat $C$, SG susceptibility $\chiSG$, and sublattice magnetization $m_{\text{s}}$ for $J_2^\mathrm{coop}=0$ [(a)-(d)] and for $J_2^\mathrm{coop}=0.075$ [(e) and (f)] at $b=0.2$. The insets in (b) and (d) show the scaling collapses of $\chiSG$.
 The inset in (f) presents $m_{\text{s}}^2(\Delta)$ at $T=0.25, 0.2$, and 0.1 for $L=4$.
 }
 \label{fig:data}
\end{figure}

We start from the result in the absence of $J_2^\mathrm{coop}$.
A typical phase diagram is presented in Fig.~\ref{fig:pd}(a) with $b=0.2$.
The system exhibits a nematic transition at $\Tc \simeq b$, but remains magnetically disordered down to zero $T$ at $\Delta=0$.
By introducing the disorder $\Delta$, the SG transition appears at a finite $\TSG$.
In the weakly-disordered region, i.e., $\Delta \lesssim b$, $\TSG$ is roughly proportional to $\Delta$; we call this regime the \textit{linear regime}.
A remarkable point is that $\TSG$ is largely enhanced compared to that in the bilinear limit: 
the enhancement factor is, e.g., about 5-10.
At $\Delta\simeq b$, $\TSG$ appears to merge into $\Tc$ with showing multicritical behavior.
For larger $\Delta$, $\TSG~(=\Tc)$ becomes nearly independent of $\Delta$, which we call the \textit{plateau regime}.
This is in sharp contrast to the previously-reported SG behavior, $\TSG \propto \Delta$~\cite{Saunders07, Andreanov10, Tam10}.
These peculiar dependences of $\TSG$ on $\Delta$ are universally obtained for a wide range of $0<b<J$.

Figures~\ref{fig:data}(a) and \ref{fig:data}(b) show typical MC data in the linear regime $\Delta \lesssim b$.
As seen in Fig.~\ref{fig:data}(a), $Q^2$ exhibits a steep rise at $T\simeq 0.22$, signaling the nematic transition.
At the same $T$, the specific heat $C$ exhibits a sharp peak.
By extrapolating the peak position to the bulk limit, we obtain $\Tc = 0.2245(15)$ at $\Delta=0.1$.
On the other hand, $\chiSG$ appears to diverge at a lower $T$, indicating the SG transition.
We obtained $\TSG= 0.102(14)$, $\gamma=2.24(75)$, and $\nu=1.16(18)$ by the finite-size scaling analysis as demonstrated in the inset of Fig.~\ref{fig:data}(b):
Here, we assume $\chiSG = L^{\gamma/\nu} f(L^{1/\nu}t)$, where $t=(T-\TSG)/\TSG$, $\nu$ and $\gamma$ are the critical exponents for the correlation length and $\chiSG$, respectively.
The critical exponents are consistent with those in the bilinear limit $b=0$~\cite{Saunders07, Andreanov10, Tam10} as well as of the canonical SG~\cite{CanonicalSG} within the error bars, but surprisingly, $\TSG$ is strongly enhanced by a factor of 5-10 from $\TSG=0.01$-$0.02$ at $b=0$~\cite{Saunders07, Andreanov10, Tam10}.

What happens in this linear regime is the following.
The spin collinearity emergent below $\Tc$ enforces spins to satisfy the spin-ice type local constraints and form locally-correlated collinear objects.
There, the system bears a semidiscrete degenerate manifold with a multivalley energy landscape, as in the case of $\Delta=0$.
Therefore, thermal fluctuations are strongly constrained compared to the bilinear case in which the ground-state manifold is continuously-connected.
At the same time, the spin-spin correlations are much enhanced to exhibit quasi-long-range behavior below $\Tc$~\cite{Shannon10}.
These are the origin of the striking enhancement of $\TSG$ demonstrated in Fig.~\ref{fig:pd}(a).

With further increasing $\Delta$, $\TSG$ reaches $\Tc$ at $\Delta\simeq b$ but never exceeds $\Tc$~\footnote{Almost flat behavior of $\TSG$ in this regime is presumably because of the competition between the randomness in $b_{ij}$ and $J_{ij}$: The former suppresses local spin collinearity, while the latter does the opposite~\cite{Bellier-Castella01}.}.
This is understood by the above physical picture. 
Typical data in the plateau regime are shown in Figs.~\ref{fig:data}(c) and \ref{fig:data}(d). 
Following the above way, we obtained $\Tc=0.256(3)$ and $\TSG=0.252(4)$ at $\Delta=0.5$.
The scaling analysis of $\chiSG$ is compatible with the second order transition with critical exponents $\gamma=1.2(5)$ and $\nu=0.67(16)$ [see the inset of Fig.~\ref{fig:data}(d)], which are also consistent with those in the bilinear limit~\footnote{Slight discrepancy in $\nu$ between $\Delta=0.1$ and $\Delta=0.5$ might be due to finite-size effects under the influence of crossover from the weak first-order transition.}.
These results indicate that the SG transition coincides with the onset of spin collinearity, and the concomitant transition looks second order in this plateau regime.
The critical behavior is indistinguishable from the conventional SG~\cite{Saunders07, Andreanov10, Tam10, CanonicalSG}: 
It is noteworthy that the peak of $C$ is suppressed and broadened [Fig.~\ref{fig:data}(c)], as seen in the canonical SG transition~\cite{CanonicalSG}.

Next, we examine the effect of the cooperative coupling $J_2^\mathrm{coop}$. 
A typical phase diagram is shown in Fig.~\ref{fig:pd}(b) with $J_2^\mathrm{coop}=0.075$ and $b=0.2$. 
At this value of $J_2^\mathrm{coop}$, the nematic phase in the linear regime is 
taken over by the ${\bf q}={\bf 0}$ N\'{e}el-ordered phase. 
Figures \ref{fig:data}(e) and \ref{fig:data}(f) indicate a first-order transition at $\TN=0.291(3)$ for $\Delta=0.2$. 
On the contrary, the SG in the plateau regime remains robust; we confirmed that the MC results in this regime are essentially the same as at $J_2^\mathrm{coop}=0$ shown in Figs.~\ref{fig:data}(c) and \ref{fig:data}(d).
Thus, $J_2^\mathrm{coop}$ lifts the residual semi-discrete degeneracy in the nematic phase, hardly affecting the SG in the plateau regime. This results in a bicritical-like phase competition between the N\'{e}el and SG phases.

The robust plateau behavior of $\TSG$ at a largely enhanced value gives an explanation for the puzzling behaviors in the pyrochlore-based antiferromagnets mentioned in the introduction. 
For example, in (La$_x$Y$_{1-x}$)$_2$Mo$_2$O$_7$, 
many experiments suggest a substantial bond disorder even in the stoichiometric case at $x=0$~\cite{Booth00,Keren01,Sagi05,Greedan09}. 
Suppose that the system inevitably includes a substantial disorder even in the best-quality sample ever made, as suggested by the microscopic probes~\cite{Booth00, Ofer10}, 
and is already in the plateau regime, the system does not pass through the nematic phase and $\TSG$ can be large and remain almost constant against the additional disorder by La substitution $x$.
In the case of (Zn$_{1-x}$Cd$_x$)Cr$_2$O$_4$, the results at a finite $J_2^\mathrm{coop}$ give a qualitatively reasonable explanation for the phase competition between the spin-lattice ordered phase and SG phase as well as the peculiar behavior of $\TSG$.
For further qualitative comparison, more sophisticated modeling is indispensable for farther neighbor exchanges, substitution effect, and spin-lattice couplings.

\begin{figure}[t]
 \centering
 \includegraphics[width=0.5\textwidth,clip]{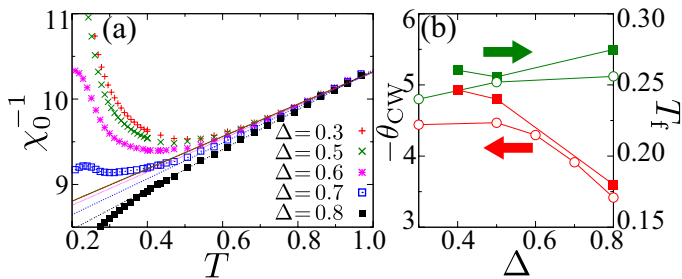}
 \caption{(color online).
(a) Inverse magnetic susceptibility $\chi_0^{-1}$ at $J_2^\mathrm{coop}=0$ for $L=3$.
The lines denote Curie-Weiss fittings for $0.6 \le T\le 0.9$.
(b) $\Delta$ dependences of $\theta_\mathrm{CW}$ and $\TSG$. The circles and squares denote the results for $J_2^\mathrm{coop}=0$ and 0.075, respectively.}
 \label{fig:inv-xi}
\end{figure}

Finally, we make a further comparison with experiments by the uniform magnetic susceptibility.
The susceptibility, calculated by
$\chi_0=(3\Ns)^{-1} \langle \langle |\sum_{i=1}^{\Ns} \vec{S}_{i}|^2/T \rangle_T \rangle_\Delta$, is shown for various values of $\Delta$ with $J_2^\mathrm{coop}=0$ in Fig.~\ref{fig:inv-xi}(a).
The results indicate that $\chi_0$ show a Curie-Weiss-like behavior at much higher $T$ than $\TSG$ with a $\Delta$-dependent Curie-Weiss temperature $\theta_\mathrm{CW}$.
As shown in Fig.~\ref{fig:inv-xi}(b), $|\theta_\mathrm{CW}|$ decreases significantly with increasing $\Delta$, whereas $\TSG$ stays almost constant in the plateau regime.
This constrasting behavior of $\theta_\mathrm{CW}$ and $\TSG$ is robust for finite $J_2^\mathrm{coop}$.
The result qualitatively agrees with the peculiar SG behavior observed in (La$_x$Y$_{1-x}$)$_2$Mo$_2$O$_7$~\cite{Sato87}.
Note that similar behaviors of $\TSG$ and $\theta_\mathrm{CW}$ were reported in $R_2$Mo$_2$O$_7$ for a change of $R$ cations, such as Gd, Tb, and Dy~\cite{Katsufuji00,Sato86}.

To summarize, through the numerical analysis of the bond-disordered pyrochlore antiferromagnet, 
we have revealed that the SG transition temperature $\TSG$ is largely enhanced by the spin-lattice coupling $b$, and becomes saturated at a temperature set by $b$.
The results reproduce the puzzling characteristics of the spin-glass transition observed in the pyrochlore-based geometrically frustrated materials. 
Unfortunately, the value of $b$ is not available for the compounds that we are concerned here~\footnote{$b$ is suggested to be $\sim 0.1$ in related spinel compounds, CdCr$_2$O$_4$ and HgCr$_2$O$_4$. See Y. Motome {\it et al.}, J. Mag. Mag. Mater. {\bf 300}, 57 (2006).}, but we believe that our results stimulate experiments to estimate of $b$, e.g., from high-$T$ nonlinear magnetic susceptibility~\cite{Kobler97}.
Since the peculiar behaviors originate in the simple local physics, we believe that they are common to systems coupled to local lattice distortions.

We thank T. Kato, H. Kawamura, K. Penc, and N. Shannon for fruitful discussions. Numerical calculation was partly carried out at the Supercomputer Center, ISSP, Univ. of Tokyo. This work was supported by Grant-in-Aid for Scientific Research (No. 19052008), Global COE Program ``Physical Sciences Frontier'', the Next Generation Super Computing Project, and Nanoscience Program, from MEXT, Japan.


\end{document}